\documentclass[prl,twocolumn,superscriptaddress]{revtex4-2}   
\usepackage{graphicx,amssymb}
\usepackage{color,ulem}
\usepackage{bm}   

\usepackage{bm}% bold math
\usepackage{amssymb,amsmath}
\usepackage{amsthm}

\usepackage{latexsym}
\usepackage{graphicx,amssymb}
\usepackage{color,ulem}
\usepackage{float}
\usepackage{siunitx}
\usepackage{mathrsfs} 
\usepackage{soul}
\renewcommand{\Re}{\mathop \mathrm{Re}}

\DeclareMathOperator{\sech}{sech}

\usepackage{graphicx}% Include figure files
\usepackage{dcolumn}% Align table columns on decimal point
\usepackage{bm}% bold math
\usepackage[utf8]{inputenc}
\usepackage{amsfonts,amssymb,amsmath}
\usepackage[mathscr]{eucal}
\usepackage{epsfig}

\begin{document}
% \preprint{AIP/123-QED}

\title{Ultimate accuracy of frequency to power conversion by single-electron injection}
% \thanks{A footnote to the article title}

\author{Jukka P. Pekola}
\affiliation{Pico group, QTF Centre of Excellence, Department of Applied Physics,
    Aalto University, P.O. Box 15100, FI-00076 Aalto, Finland}
\author{Marco Mar\'in-Su\'arez}
\affiliation{Pico group, QTF Centre of Excellence, Department of Applied Physics,
	Aalto University, P.O. Box 15100, FI-00076 Aalto, Finland}
\author{Tuomas Pyh\"aranta}
\affiliation{Pico group, QTF Centre of Excellence, Department of Applied Physics,
	Aalto University, P.O. Box 15100, FI-00076 Aalto, Finland}
\author{Bayan Karimi}
\affiliation{Pico group, QTF Centre of Excellence, Department of Applied Physics,
	Aalto University, P.O. Box 15100, FI-00076 Aalto, Finland}
\affiliation{QTF Centre of Excellence, Department of Physics, Faculty of Science, University of Helsinki, FI-00014 Helsinki, Finland}

\date{\today}

\begin{abstract}
We analyze theoretically the properties of the recently introduced and experimentally demonstrated converter of frequency to power.
The system is composed of a hybrid single-electron box with normal island and superconducting lead, and the detector of the energy flow using a thermometer on a normal metal bolometer. Here we consider its potential for metrology.
The errors in power arise mainly from inaccuracy of injecting electrons at the precise energy equal to the energy gap of the superconductor.
We calculate the main systematic error in form of the excess average energy of the injected electrons and its cumulants, and due to sub-gap leakage.
We demonstrate by analytic and numerical calculations that the systematic error in detection can, in principle, be made much smaller than the injection errors, which also, with proper choice of system parameters, can be very small, $< 1$\%, at low enough temperature.
Finally we propose a simplified configuration for metrological purposes.   
 
\end{abstract}

\maketitle
\begin{figure}
	\centering
	\includegraphics [width=\columnwidth] {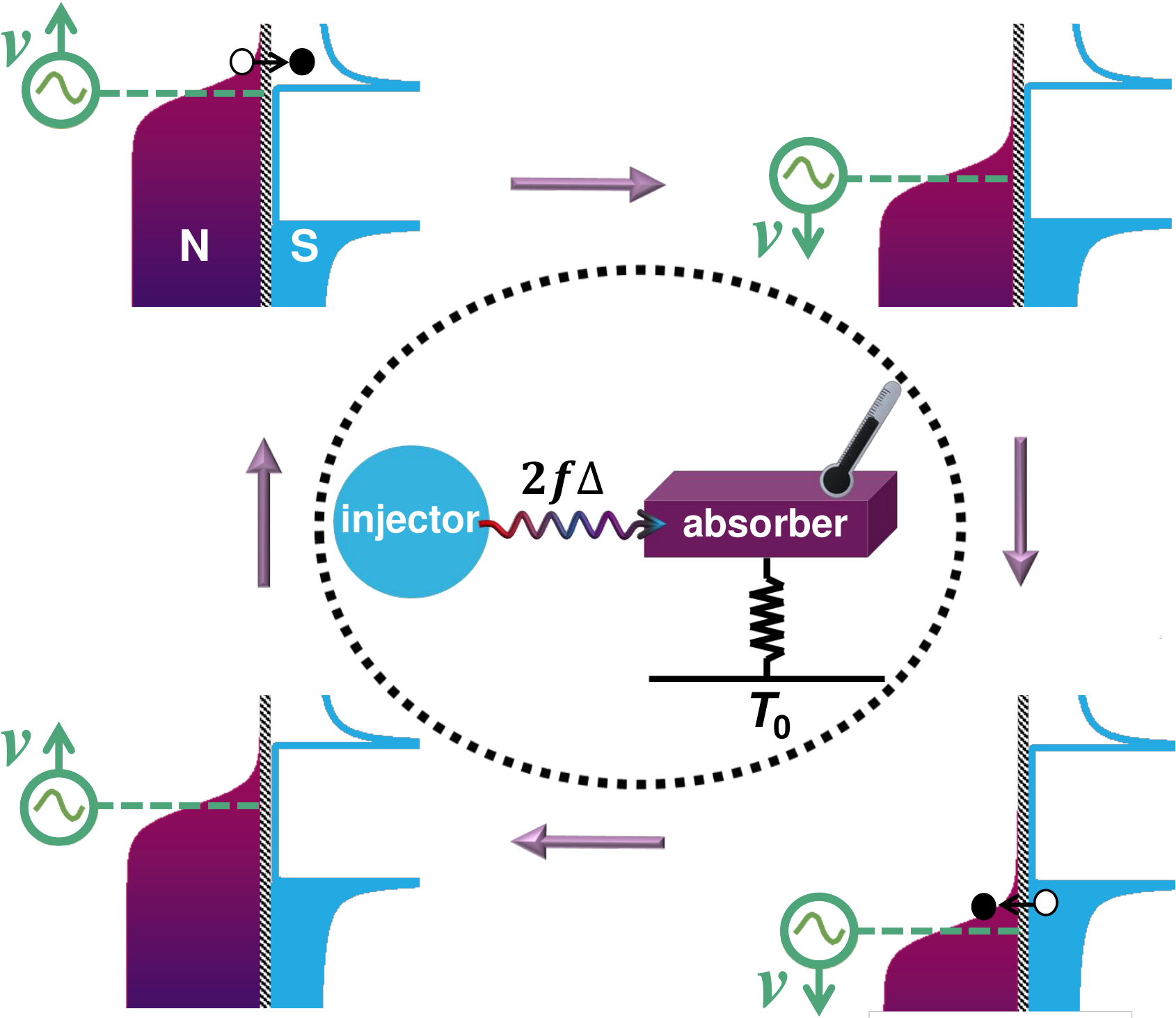}
	\caption{The pumping cycle to convert frequency to power.
	The potential of the normal (N) electron box is shifted periodically by the gate voltage with respect to that of the superconducting lead (S).
	Each time the N potential passes the gap energy in S, an electron tunnels to/from N creating an excitation with energy $\sim \Delta$.
	With sufficiently slow drive, at low temperature and with hard gap of the superconductor this energy approaches $\Delta$, as will be discussed quantitatively in the text.
	In the centre we illustrate the all-in-one system consisting of the power injector and absorber whose temperature is then measured.
		\label{fig1}}
\end{figure}

{\sl Introduction:} Single-electron and superconducting devices form important building blocks in modern electrical metrology.
Voltage from Josephson effect, resistance from quantum Hall effect and current from single-electron transport are all reliable ways to determine these quantities with high accuracy \cite{Bae,Scherer,Giblin,Krasnopolin,Ribeiro}.
Extending to other application areas of quantum electronic devices \cite{Bauerle,Carrega}, some of us have recently demonstrated a hybrid single-electron transistor as a frequency to power converter\cite{Marin}.
In this first experiment the accuracy in generating a desired level of power was still modest, on the level of about 10\%, mainly because of the non-optimized calibration of the bolometric detector. In this letter we discuss the fundamental errors of the frequency to power conversion arising from nonadiabaticity, noise, sub-gap leakage and temperature.
Similarly we analyze the error of the bolometric detector in form of trapping efficiency of the absorber whose temperature is monitored.
The injector and detector form an all-in-one system compatible for integrated design and fabrication, schematically depicted in the centre of Fig. \ref{fig1}.
We obtain analytical results that allow for direct assessment of experimental opportunities, and propose a simplified setup for future precision experiments.

{\sl Description of the system:} %Figure x1 presents the basic setup as employed in Ref. \cite{Marin}, converting frequency to power and detecting it. 
The heart of the system is a single-electron transistor (SINIS-turnstile, S for superconductor, I for insulator, N for normal metal), which allows both gate biasing at voltage $V_g$, and drain-source biasing at voltage $V$, see Fig. S1 in \cite{SM}.
The key idea is that periodically varying gate voltage, at frequency $f$, injects electrons between the S lead(s) and the N island such that each tunneling event adds an energy very close to the superconducting gap $\Delta$ to the lead.
In the first demonstration (Ref. \onlinecite{Marin}) such a turnstile geometry was employed allowing full characterization of the device with electrical transport measurements.
The injector works both at non-vanishing bias $V$ as well as at $V=0$; in the former case it produces ideally power $f\Delta$ to both leads, even if the tunnel contacts are not identical, whereas at $V=0$ only the total power to the two leads equals $2f\Delta $.

We reduce the device in our analysis below into a N box emitting excitations to one S lead, see Fig. \ref{fig1}.
It bears no difference in operation with respect to the turnstile configuration at zero-bias, except that all the power $P \approx 2f\Delta $ is now injected into one S lead.
This is also the physical setup that we propose for future precision experiments.
We denote the time-dependent chemical potential of the N island with respect to the S lead by $v$ (in units of $\Delta$), modulated by the gate voltage.
The given result $P \approx 2f\Delta$ is easy to understand by noting that the  Bardeen-Cooper-Schrieffer (BCS) density of states (DOS) in the superconductor vanishes at sub-gap energies and has a singularity at the gap.
Therefore, under the gate drive with not too high frequency, the electrons tunnel very close to the gap energy of the superconductor.
The cycle leading to the given value of power is illustrated by the cartoon in Fig. \ref{fig1}. Systematic errors in injection energy, and thus average power over many driving periods, arise from the delayed on-demand tunneling events if the gate voltage passes the singularity condition too fast, and from fluctuations of the same effect.
In the analysis below we also take into account the non-vanishing temperature $T$, and non-idealities in tunneling. We parametrize the latter effect by the smearing of the BCS DOS using the commonly adopted Dynes model \cite{Dynes1,Dynes2}, which can be directly related to experimental parameters \cite{jltp2022}. 

Besides injection, errors in $f$ to $P$ conversion can incur due to the detection of $P$, which is done bolometrically.
The excitations, electron- or hole-like quasiparticles in the superconductor, diffuse through the superconducting wire into a normal metal absorber, and its power is measured by detecting the steady-state rise of the temperature of this N absorber \cite{Wellstood}.
Fundamental errors arise due to loss of quasiparticle energy to phonons \cite{Ullom} or via the leakage of heat through the whole SNS chain around the absorber in the diffusion process \cite{Ullom2}, and due to error of the temperature measurement which includes also the natural fluctuations of heat between electrons and phonons in the N absorber.

{\sl Analysis of the injector:} We start by assessing the injector performance in terms of deviations from the ideal $P=2f\Delta $ power.
Since, as it is to be expected, the optimal operation of the frequency to power conversion is achieved at low temperature ($k_{\rm B}T\ll \Delta$), at moderately low frequencies, and with nearly ideal junctions. In this case most of the results of interest can be obtained analytically.
This leads to quite transparent presentation with illustrative and characteristic results demonstrating scaling in terms of device parameters, driving frequency and temperature. Eventually we present both analytic and numeric results with good consistency.
The numerical analysis follows closely that presented in Ref. \onlinecite{Marin} but extended to higher moments of injected energy, see Supplementary Material \cite{SM}. 

\begin{figure*}
	\centering
	\includegraphics [width=\textwidth] {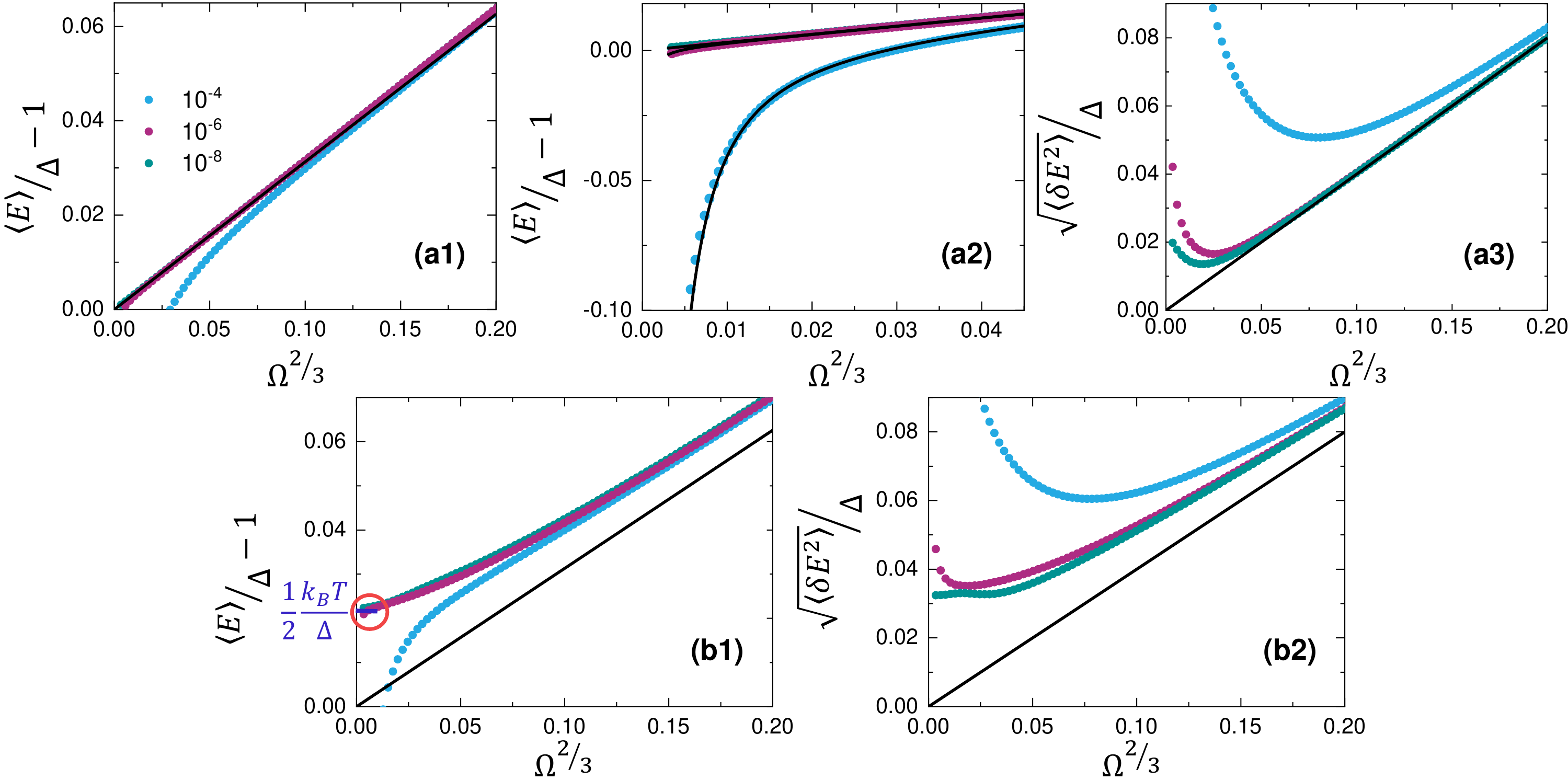}
	\caption{Main results of the numerical and analytical calculations.
	In (a), the results at zero temperature are shown against the dimensionless ramp rate $\Omega$.
	(a1) The average extra energy with respect to $\Delta$.
	The dependence $0.313 \Omega^{2/3}$ from Eq. \eqref{mean-E} (solid line) closely follows the numerical result.
	Legend gives the values of the Dynes leakage parameter $\gamma$.
	%Different sets of symbols correspond to Dynes leakage parameter values $\gamma=10^{-4},10^{-6}$, and $10^{-8}$.
	(a2) Zoom-in of the low $\Omega$ regime of (a1), together with analytical predictions of Eq. \eqref{mean-Egamma} shown by solid lines.
	(a3) Root-mean-square value of the deposited energy presented as in (a1) with the solid line from Eq. \eqref{SD_E}.
	(b) Results at non-vanishing temperature, chosen to be $T=100$ mK with aluminium, $\Delta /k_B = 2.3$ K, for practical comparison.
	(b1), (b2) Average excess energy and its root-mean-square value.
	The labels of the symbols and lines are as in (a1) and (a3).
	The circle at $\Omega=0$ in (b1) points to the equipartition result of Eq. \eqref{temperature}, in full agreement with the numerics for lowest $\gamma$ as $\Omega\rightarrow 0$.
		\label{fig2}}
\end{figure*}

%{\sl Description of the numerics of the injector:} Here, we use the numerical calculations described in the section S4 of the supplement of Ref. \onlinecite{Marin} for the turnstile operation of the SINIS transistor. To calculate the higher moments of the transmitted energy $\left\langle E^n\right\rangle$ we simply modify the vector $\mb{b}$ so that $b_i=\dot{Q}^{(n)}_{i\rightarrow i+1}+\dot{Q}^{(n)}_{i\rightarrow i-1}$, where

%\begin{equation}
%\dot{Q}^{(n)}_{i\rightarrow i+1}=\dfrac{\Delta^{n+1}}{e^2R_\mr{T}}\int{d\epsilon\epsilon^n n_\mr{S}\left(\epsilon\right)f_\mr{N}\left(\epsilon -v\right)\left[1-f_\mr{S}\left(\epsilon\right)\right]}.
%\label{stationary_moment}
%\end{equation}

{\sl Analytical description:} For small errors we make first the following Markovian assumption: the deposited energy in a pumping cycle (one event in each half-period, see Fig. \ref{fig1}) is assumed to be independent of the history.
This is justified by the fact that, with high accuracy, the system is reset to the desired charge state after each half-period.
We also assume that back-tunneling is sufficiently weak, meaning that only one favorable tunnel event occurs near the matching condition of island potential and the gap in the lead.
Therefore we obtain the average power and its cumulants by focusing on the rising part of $v(t)$, and solve the master equation $dp/dt=-\Gamma (t) p(t)$, where $p(t)$  is the survival probability of the charge on the island.
This quantity has the formal solution
\begin{equation} \label{e0}
	p(t)=p(0)e^{-\int_0^t d\tau \Gamma(\tau)}.
\end{equation}
The rate out when the island is gate-biased at $v$ reads
\begin{equation} \label{e1}
	\Gamma=\frac{\Delta}{e^2R_{\rm T}}\int d\epsilon n_{\rm S}(\epsilon)f_{\rm N}(\epsilon-v)[1-f_{\rm S}(\epsilon)].
\end{equation}
Here $R_{\rm T}$ is the tunnel junction (normal state) resistance, $f_{\rm X}(\epsilon)=1/(1+e^{\beta_{\rm X}\epsilon})$ is the Fermi-Dirac distribution of electrons on the $\rm X =\rm N,\rm S$ electrode at inverse $\beta_{\rm X} =\Delta/(k_BT_{\rm X})$ of the temperature $T_{\rm X}$, and $n_{\rm S}(\epsilon)$ is the BCS DOS in the superconductor \cite{Bardeen2}.
The energies are here normalized by $\Delta$, so that $n_{\rm S}(\epsilon)=0$, when $|\epsilon| \le 1$, and $|\epsilon|/\sqrt{\epsilon^2-1}$ otherwise.
    
With no sub-gap tunneling, we have $\Gamma=0$ at $v<1$.
At gate-biases above the gap ($v>1$), we can obtain the zero temperature rate from Eq. \eqref{e1} as
%\begin{equation} \label{e6}
	$\Gamma = \frac{\Delta}{e^2R_{\rm T}}\sqrt{v^2-1}$.
%\end{equation}
We assume a linear ramp $v =\dot v t$ in time $t$, where $\dot v$ is the constant rate of change of the chemical potential in $\Delta$ units.
Such a choice is generally valid since tunneling mainly occurs around the singularity of the BCS DOS, i.e. when $|v|\approx 1$.
This $\dot v$ is proportional to the frequency $f$ for a given waveform of pumping.
With no sub-gap tunneling, we have $p=1$ at $v\le 1$.
At gate-biases above (but close to) the gap ($v\gtrsim 1$), we obtain the zero temperature survival probability with the help of Eq. \eqref{e0} as 
\begin{eqnarray} \label{e7}
	p(v)&&=\exp [-\frac{1}{2\Omega}(v\sqrt{v^2-1}-\ln (v+\sqrt{v^2-1}))]\nonumber \\ &&\approx \exp [-\frac{2\sqrt{2}}{3}\Omega^{-1}(v-1)^{3/2}].
\end{eqnarray}
Here, $\Omega\equiv e^2R_{\rm T}\dot{v}/\Delta$ is the dimensionless frequency, which depends also on the system parameters $R_{\rm T}$ and $\Delta$.
To find the statistics of heat in this configuration we write the probability distribution as 
\begin{equation} \label{e26}
	P(\epsilon,v)=-p'(v)\Pi(\epsilon,v).
\end{equation}
Here $\Pi(\epsilon,v)$ is the probability (density) of tunneling to the state with energy $\epsilon$ in S, provided it happens at voltage $v$, given by  
%\begin{equation} \label{e27}
$	\Pi(\epsilon,v)=\pi(\epsilon,v)/\int \pi(\epsilon',v) d\epsilon'$,
%\end{equation}
where $\pi(\epsilon,v)= n_{\rm S}(\epsilon)f_{\rm N}(\epsilon-v)[1-f_{\rm S}(\epsilon)]$ is the spectral rate at energy $\epsilon$ when biased at $v$.
We see immediately that $P(\epsilon,v)$ is properly normalized, i.e. $\int P(\epsilon,v) d\epsilon dv =1$.
We then obtain the moments of the transmitted energy $\langle E^n\rangle$ as
\begin{equation}\label{E^n}
	\langle E^n\rangle=-\Delta^n\int dv~p'(v) \int d\epsilon~\epsilon^n~\Pi(\epsilon,v).
	\end{equation}
We analyze the lowest moments at zero temperature analytically as a power series of $\Omega$. In the lowest order we obtain %In particular, for the cumulants $\langle \delta E^n\rangle\equiv \langle (E-\langle E\rangle)^n\rangle$ we obtain in the lowest order
\begin{equation}\label{mean-E}
		\frac{\langle E\rangle}{\Delta}=1+\frac{\Gamma(2/3)}{3^{4/3}}\Omega^{2/3}\approx 1+0.313\,\Omega^{2/3}.
\end{equation}
Naturally the average energy exceeds $\Delta$ since the minimum energy that the tunneling can bring equals $\Delta$.
We also calculate the cumulants $\langle \delta E^n\rangle\equiv \langle (E-\langle E\rangle)^n\rangle$.
In particular, the standard deviation of the deposited energy is given by
	\begin{eqnarray}\label{SD_E}
		\frac{\sqrt{\langle \delta E^2\rangle}}{\Delta}&&=(\frac{3}{2\sqrt{2}})^{2/3}\sqrt{\frac{\Gamma(7/3)}{5}-\frac{\Gamma(5/3)^2}{9}}\Omega^{2/3}\nonumber \\&&\approx 0.400\,\Omega^{2/3}.
	\end{eqnarray}
Similarly for the cubic root of the third cumulant in units of $\Delta$, we have
	\begin{eqnarray}\label{third-E}
		\frac{\sqrt[3]{\langle \delta E^3 \rangle}}{\Delta}&&=\sqrt[3]{\frac{9}{28}+\frac{1}{12}\Gamma(5/3)^3-\frac{9}{40}\Gamma(5/3)\Gamma(7/3)}\Omega^{2/3}\nonumber\\&&\approx 0.520\,\Omega^{2/3}.
	\end{eqnarray}
We note that all the quantities in Eqs. \eqref{mean-E} -- \eqref{third-E} scale as $\Omega^{2/3}$.
We also see that, as could be expected from the strong asymmetry of the DOS around $\epsilon = \Delta$, the average energy in Eq. \eqref{mean-E} exceeds $\Delta$, and the skewness based on Eq. \eqref{third-E} is always positive. 
%We will return to the implications of these results after we have analyzed the whole error chain including numerical results and the analysis of the bolometer efficiency.

{\sl Influence of the sub-gap leakage and non-vanishing temperature:} 
Besides the influence of driving frequency, deviations from the ideal $P=2f\Delta$ arise from temperature-independent subgap leakage and from non-vanishing temperature that were ignored above. For the first one, we assume the standard Dynes form of the DOS $n_{\rm S}(\epsilon)=\left|\Re \left[(\epsilon+i\gamma)/\sqrt{(\epsilon+i\gamma)^2-1}\right]\right|$, with the dimensionless smearing parameter $\gamma$. To obtain the average heat in a half-period, we follow the same procedure as before, but now with the sub-gap DOS $n_{\rm S}(\epsilon) \approx \gamma/(1-\epsilon^2)^{3/2}$ for $|\epsilon| < 1$ and subsequent survival probability $p(1)=e^{-\gamma/\Omega}$. With these we obtain the zero-temperature average heat as
\begin{equation}\label{mean-Egamma}
	\frac{\langle E\rangle}{\Delta}\approx (\frac{\pi}{2}-1)\frac{\gamma}{\Omega}e^{-\gamma/\Omega}+ (1+0.313\,\Omega^{2/3})e^{-\gamma/\Omega},
\end{equation} 
which coincides with Eq. \eqref{mean-E} for $\gamma \rightarrow 0$. Naturally the influence of non-vanishing $\gamma$ is to lower $\langle E\rangle$ due to tunneling events below the gap. %The analytic expression of Eq. \eqref{mean-Egamma} is shown together with the numerical results in Fig. xx.

For the non-vanishing temperature, we return to the ideal BCS DOS to obtain the leading error at low $\Omega$ and vanishing $\gamma$. Again with the same procedure as above, we find the survival probability at $v=1$ as $p(1)\approx e^{-\sqrt{\pi/2}\,\Omega^{-1}(k_BT/\Delta)^{3/2}}$. Therefore,  for $\Omega\ll \sqrt{\pi/2}\,(k_BT/\Delta)^{3/2}$, the tunneling mainly occurs already at $v<1$ but at energy just above the gap due to the tail in the Fermi-Dirac distribution in N. With standard integrations, this yields the the result at $\Omega \rightarrow 0$ as
\begin{equation} \label{temperature}
\frac{\langle E\rangle}{\Delta}\approx 1+ \frac{1}{2}\frac{k_BT}{\Delta},\,\,\, \Omega,\gamma \rightarrow 0,
\end{equation}
presenting an equipartition-like excess energy in this quasistatic process.

%Next we wish to compare the results of numerical simulations to the obtained analytical ones. 
Figure \ref{fig2} presents the main characteristics of the injector performance. Since both $\langle E\rangle-\Delta$ and $\sqrt{\langle \delta E^2\rangle}$ are proportional to $\Omega^{2/3}$ according to the analytic expressions \eqref{mean-E} and \eqref{SD_E}, we use this as the horizontal scaling. We observe that the numerical results for the two quantities agree quantitatively with the predictions \eqref{mean-E} and \eqref{SD_E} for $T=0$ and the smallest $\gamma$ ($=10^{-8}$). For non-vanishing leakage, in particular for  
$\gamma=10^{-4}$, Eq. \eqref{mean-Egamma} yields a decent approximation which deviates from the linear behavior especially at low values of $\Omega$ (see Fig. \ref{fig2} (a2)). Likewise, the prediction of Eq. \eqref{temperature} captures the quasistatic excess energy towards $\Omega=0$ at non-vanishing temperatures. Based on these successful comparisons, we conclude that Eqs. \eqref{mean-E} - \eqref{temperature} form a perfect basis of assessing the errors in frequency-to-power conversion.

\begin{figure}
	\centering
	\includegraphics [width=\columnwidth] {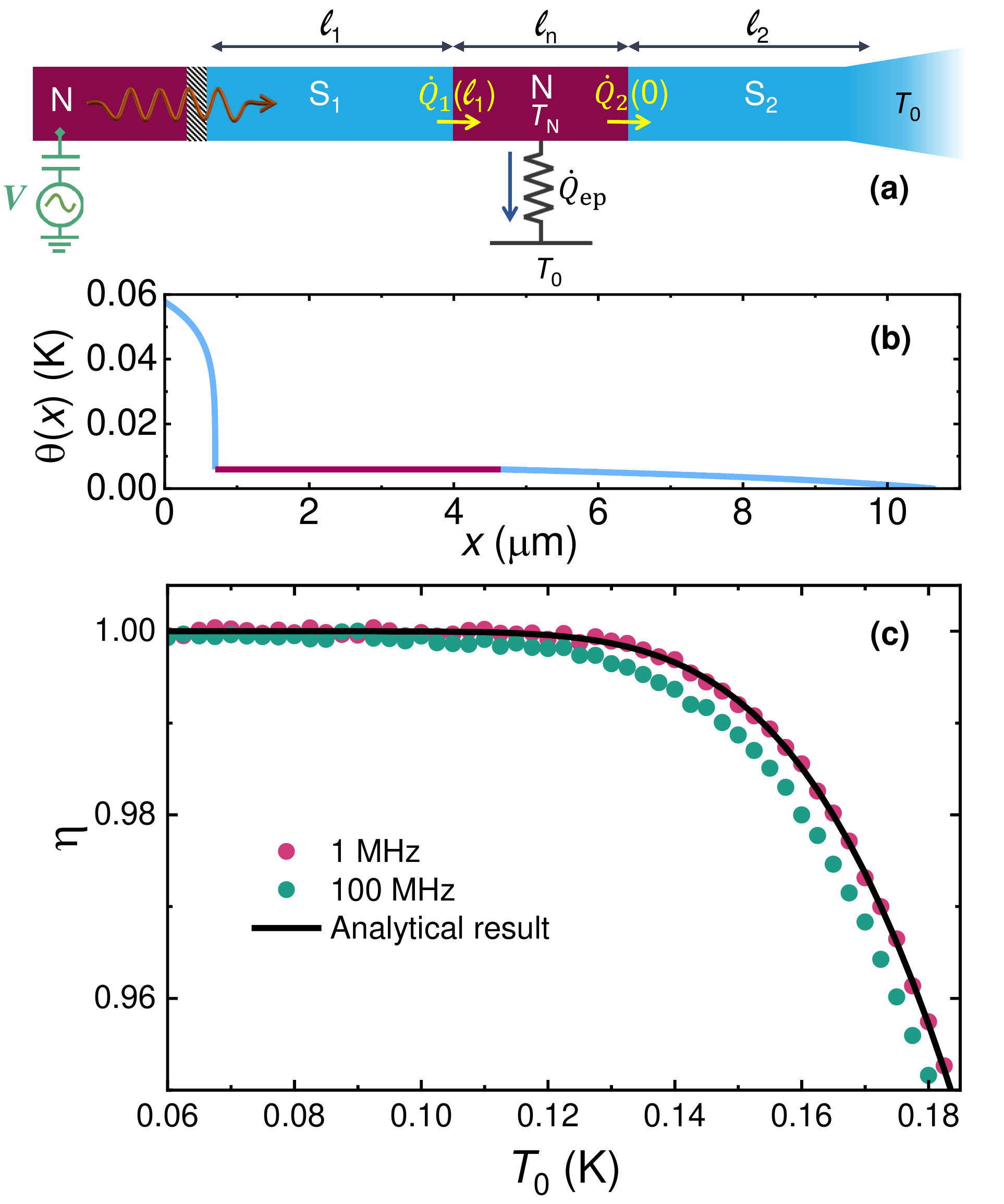}
	\caption{The bolometric detector. (a) The scheme. The power injected to the superconductor S1 diffuses into the normal absorber N and leaks partly to the second superconductor S2 which is thermalized to bath temperature $T_0$ at the far end. (b) Temperature profile along the structure with coordinate $x$ in (a) with the following parameters: $\dot{Q}$ corresponds to $f=10\,{\rm MHz}$ and $T_0=100\,{\rm mK}$, the geometric and material parameters can be found in \cite{SM}. (c) Trapping efficiency $\eta$ as a function of $T_0$. The device parameters are as in (b). The numerical results shown by solid symbols are calculated with two injection frequencies. The line is from Eq. \eqref{etaappr}.
		\label{fig4}}
\end{figure}

{\sl Analysis of the bolometer efficiency:}
In the structure of Fig. \ref{fig4} (a), the lead into which the excitations are injected is composed of three sections S1 (superconductor), N (normal metal) and S2 (superconductor), connected by clean metallic contacts to each other.
The section lengths and cross-sectional areas are, respectively, denoted by $\ell_i$ and $A_i$ with proper subscripts, and the volume of the N absorber is $\mathcal V_{\rm N}$.
By combining the continuity equation with Fourier's law, we can derive a one-dimensional model for heat transport.
In this model, the temperature difference obeys a nonlinear differential equation of the form 
\begin{align}
	\frac{d}{dx} \left[ - \kappa(x) A \frac{d}{dx} \theta(x) \right] - \dot Q_{\rm ep}(x) = 0,
\end{align}
in each of the three regions, with $\theta(x)=T(x)-T_0$ the local temperature with respect to that of the bath, $T_0$.
The problem satisfies the following boundary conditions: (i) heat current into the S1 wire on the left end equals $\dot Q$ (approximately $2f\Delta$ in practice), (ii) the temperature is continuous across the SN interfaces, (iii) similarly, the heat current is continuous in these interfaces, and (iv) temperature at the right end of S2 equals $T_0$ since this end is intentionally thermalized to the phonon temperature by a big reservoir.
Both the thermal conductivity $\kappa(x)$ as well as the power transfer per unit length due to the the electron-phonon interaction $\dot Q_\mathrm{ep}(x)$ depend on temperature.
For the superconducting regions $\kappa_{\rm S} = \frac{2\Delta^2}{e^2\rho T_{\rm S}}e^{-\Delta/k_BT_{\rm S}}$ \cite{Bardeen} and $\dot Q_{\rm ep, S}=a\Sigma_{\rm S}  A_{\rm S} T_0^4 e^{-\Delta/k_BT_{\rm S}}\theta(x)$ as derived from \cite{Maisi}.
Here $\rho$ is the normal state resistivity of the superconductor, $\Sigma_{\rm S}$ is the material specific electron-phonon coupling parameter, and $a \sim 5$ is a numerical constant.
For the normal-metal the heat conductance is given by the Wiedemann-Franz law and $\dot Q_{\rm ep, N}=\Sigma_{\rm N}A_{\rm N}\left(T^5_{\rm N}(x)-T_0^5\right)$ with $\Sigma_{\rm N}$ the electron-phonon parameter for the normal metal \cite{Wellstood}.

The resulting boundary value problem can be solved numerically using the shooting method, which is most conveniently applied using the heat power at the end of S2, $\dot{Q}_\mathrm{2}(\ell_2)$, as the shooting parameter as it is bounded from above and below by $2f\Delta$ and zero, respectively.
One then varies $\dot{Q}_2(\ell_2)$ until the boundary conditions are satisfied, i.e. until the power at the injection point matches $2f \Delta$.
Figure \ref{fig4} (b) depicts a typical solution of the temperature difference along the SNS wire. 

{\sl Linearization:} For small temperature differences, $\theta (x)/T_0\ll 1$, valid for low-frequency operation or high bath temperatures, we may neglect the nonlinearities, and we have approximately
\begin{equation} \label{lintheta}
	\theta''(x)-\lambda^2 \theta(x)=0,
\end{equation}
where $\lambda=\sqrt{a\Sigma_{\rm S}e^2\rho/(2\Delta^2)} T_0^{5/2}$.
In sections S1 and S2 we solve Eq. \eqref{lintheta} whereas in N we assume a constant temperature $T_{\rm N}$ throughout because of the good thermal conductivity in it (see the normal-metal section of plot in Fig. \ref{fig4} (b)).
This procedure allows one to solve analytically the temperature profile in the SNS wire for a given set of parameters including $\dot Q$, $T_0$ and the geometric and material parameters. Our prime interest is to find the trapping efficiency $\eta$ of the bolometer, which is the ratio of the power absorbed in N and the injected power, i.e. $\eta= \dot Q_{\rm N}/\dot Q$ with $\dot Q_{\rm N}=\dot Q_1(\ell_1)-\dot Q_2(0)$, the difference of incoming and outcoming heat fluxes at the ends of the two superconductors at the intersection with N, see Fig. \ref{fig4} (a). We find then in this linearized model
\begin{equation} \label{etafull}
	\eta=\frac{\sech \lambda \ell_1}{1+\varpi(A_1 \tanh \lambda \ell_1+A_2 \coth \lambda \ell_2)T_0^{-5}e^{-\Delta/k_B T_0}}. 
\end{equation}
with $\varpi = \frac{2\lambda\Delta^2}{5e^2\rho \Sigma_{\rm N}\mathcal V_ {\rm N}}$. This is the general expression in the linear regime for arbitrary length of the wire. Note that the expression does not depend on the input power. To simplify Eq. \eqref{etafull}, we find for aluminium superconductor at $T_0 = 100 $ mK that $\lambda \sim 10^2 $ m$^{-1}$, meaning that the phonon relaxation length is about 1 cm. Then for the typical structure with $\ell_i\ll 100\,\mu{\rm m}$, $\lambda\ell_i \ll 1$, and Eq. \eqref{etafull} can be approximated to a good accuracy by
\begin{equation} \label{etaappr}
	\eta\approx (1+\frac{\varpi A_2}{\lambda \ell_2}T_0^{-5}e^{-\Delta/k_B T_0})^{-1}.
\end{equation}

A comparison between the numerical results and the prediction of Eq.~\eqref{etaappr} is presented in Fig. \ref{fig4} (c).
We can see that the differences are negligible especially for small frequencies.

\begin{figure}[h!]
	\centering
	\includegraphics [width=\columnwidth] {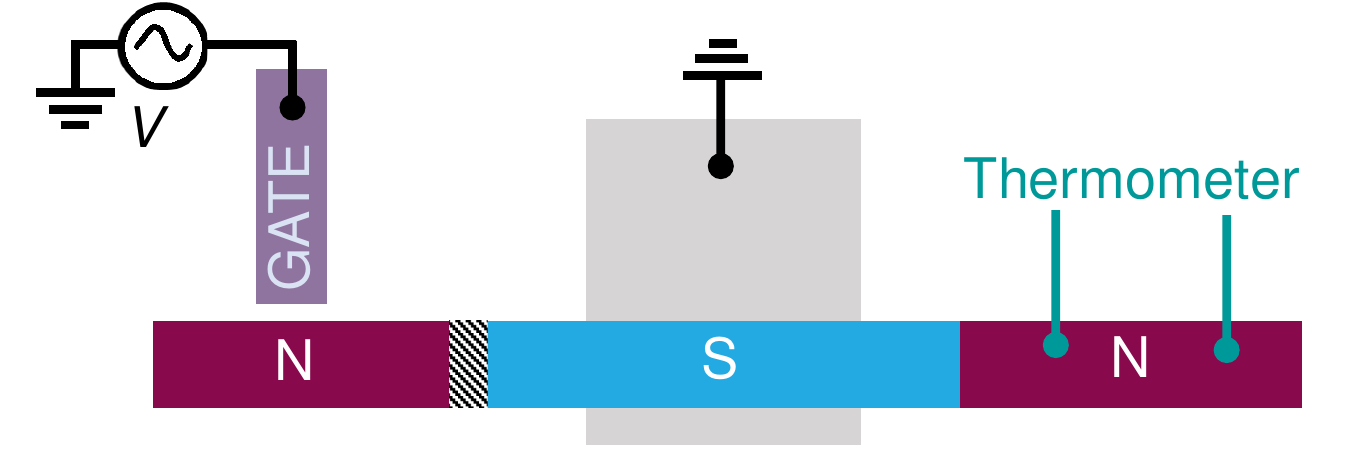}
	\caption{Proposed improved configuration of the power source. The superconductor S2 is removed from the original configuration and the potential of the S lead is kept fixed by a large capacitance to ground.     
		\label{fig5}}
\end{figure}

Equation \eqref{etafull} demonstrates that the main source of reduced efficiency is the leakage of heat through superconductor S2. In fact there are ways to suppress this effect substantially: (i) if one wishes to employ the configuration as in the experiment of Ref. \onlinecite{Marin} in order to be able to characterize \textit{in-situ} the parameters of the injector, one may still increase the length $\ell_2$, which according to Eq. \eqref{etaappr} decreases this leakage-induced loss efficiently. (ii) If one can sacrifice the transport characterization, one may improve the setup further by simplifying it into what is sketched in Fig. \ref{fig5}, i.e. one may drop out S2 completely. This way the efficiency simplifies into 
\begin{equation} \label{etaappr2}
	\eta\approx \sech{ \lambda\ell_1},
\end{equation} 
determined then by the electron-phonon leakage in S1. In this configuration $\eta$ drops only $0.005\%$ from unity for temperatures around $100\,{\rm mK}$ and $\ell_1 \lesssim 100\,{\rm \mu m}$, and exponentially less at lower temperatures. In order to secure proper gate modulation, one may prefer to shunt the S1 lead in this case by a large capacitance to ground.
{\sl Discussion and conclusions}: 
Finally we make some practical estimates based on standard materials and experimental constraints. Before that we note that the systematic error in $\langle E\rangle$ is the same as that in average power, whereas the standard deviation in power diminishes as $\sqrt{\langle \delta E^2\rangle/n}$, where $n$ is the number of pumping cycles averaged in measuring the power. Typically $n$ is large, $n > 10^6$ for a slow temperature measurement, and therefore the noise in injection can be essentially neglected in comparison to the systematic errors in $P$. For high quality aluminium-based junctions, $\gamma$ can be as small as $<10^{-6}$ \cite{gamma}, which makes its influence quite minor at practical frequencies of operation. Apparently the temperature error of Eq. \eqref{temperature} is the most fundamental one and hardest to suppress. Using aluminium as the superconductor, $\Delta/k_B \approx 2.3$ K for thin films, and, according to Eq. \eqref{temperature}, the error in $\langle E\rangle$ can be made $< 1$\% if the temperature is $T \lesssim 50$ mK. At the same time one needs to keep the driving frequency $f < 100$ MHz based on Eq. \eqref{mean-E}, this assuming $R_{\rm T}=30$ k$\Omega$ and $\dot v \sim 2f$, the exact prefactor of the latter depending on the waveform, and proportional to driving amplitude and island charging energy. For even smaller errors, one needs to consider lower temperatures or alternative superconductors with larger gap, but from the point of low $\gamma$, aluminium is the choice. Practical errors in the bolometric detection are two-fold. The ones related to efficiency of the N trap at low $T$ vanish as $1-\eta \propto T^{-5}e^{-\Delta/k_BT}$ according to Eq. \eqref{etaappr}, meaning that this error is $\ll 1$\% even at $T<150$ mK, and thus fully manageable. As a last point we note that the fundamental noise in measuring the temperature of the absorber is dictated by the thermal electron-phonon noise given near equilibrium by the fluctuation-dissipation theorem. The heat current noise $S_{\dot Q} = 10\Sigma_{\rm N} \mathcal V k_BT^6$, and the rms noise with measurement bandwidth $\nu$ is $\sqrt{\langle \dot Q^2\rangle} =\sqrt{S_{\dot Q}\nu}$. Signal-to-noise ratio SNR is then SNR $=2f\Delta/\sqrt{\langle \dot Q^2\rangle}$. With typical parameters, this yields SNR $\approx 10^5$ for $f=100$ MHz, $\nu=1$ Hz and $T=100$ mK, and thus SNR $\gg 1$ for all reasonable values of parameters. Therefore the accuracy of the frequency-to-power conversion is not limited fundamentally by the bolometric detection. 

In summary, the main fundamental errors in frequency to power conversion stem from non-vanishing temperature $\left(\frac{1}{2}k_BT/\Delta\right)$, and from non-vanishing operation frequency (in a real setup $\propto \left(R_{\rm T}f/\Delta\right)^{2/3}$). The junction quality via the subgap leakage poses another limitation with error vanishing exponentially with the "hardness" of the gap. Here we have analyzed a BCS-superconductor as the energy filter; similar analysis could be done for other types of emitters such as single-level quantum dots. Finally we believe that the present error analysis can possibly be complemented by pumping error accounting \cite{Reifert}.

\section{Acknowledgements:} We thank Dmitry Golubev for many useful discussions. This work was supported by Academy of Finland (grant number 312057).

\end{document}

\title{Supporting Information\\ An electron turnstile for frequency to power conversion}
\author{Jukka P. Pekola}\email{jukka.pekola@aalto.fi}
\author{Marco Marín-Suárez}
\author{Tuomas Pyhäranta}
\affiliation{Pico group, QTF Centre of Excellence, Department of Applied Physics, Aalto University, FI-000 76 Aalto, Finland}
\author{Bayan Karimi}
\affiliation{Pico group, QTF Centre of Excellence, Department of Applied Physics, Aalto University, FI-000 76 Aalto, Finland}
\affiliation{QTF Centre of Excellence, Department of Physics, Faculty of Science, University of Helsinki, FI-00014 Helsinki, Finland}

\maketitle

\section{Initial proposal}

\begin{figure}[ht!]
\includegraphics[scale=0.5]{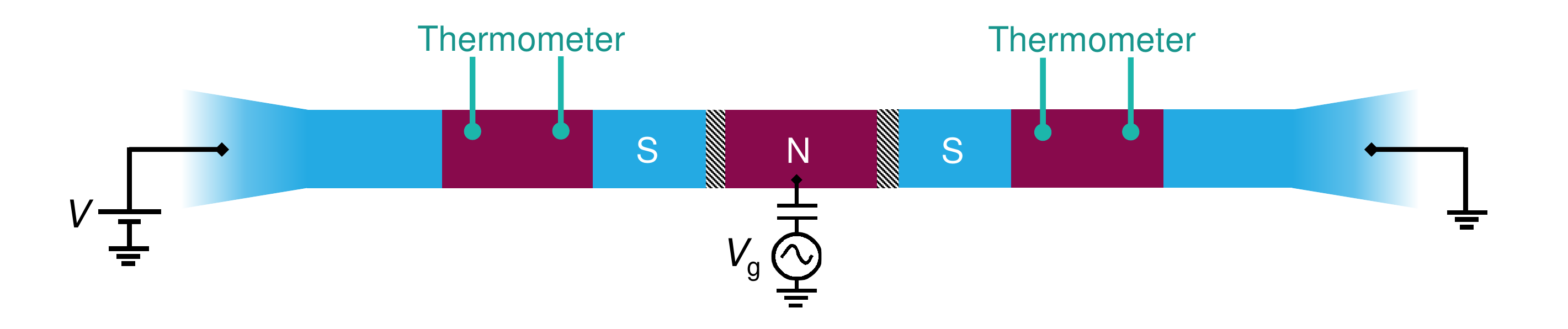}
\caption{Experimental setup used in Ref. \onlinecite{Marin-Suarez2022}. A SINIS transistor (light-blue indicates superconductor, purple indicates normal-metal) is driven by a gate voltage $V_\mr g$ via a capacitively coupled gate electrode, while at the same time biased by a DC voltage $V$. The leads of the transistor are directly contacted by normal-metals whose temperature is monitored via thermometers and then transduced to the power injected to them.}
\label{setup}
\end{figure}

The first proposal of frequency-to-power conversion made in Ref. \onlinecite{Marin-Suarez2022} consists of a SINIS (S for superconductor, I for insulator and N for normal-metal) single-electron transistor, whose leads are clean-contacted to normal-metal traps that act as bolometers for the energy carried by the injected quasiparticles (see Fig. \ref{setup}). In turn, these traps are directly interfaced by long superconductors allowing for drain-source biasing and electrical characterization.

\section{Description of the numerics of the injector}\label{S4}

Starting from the Markovian assumption, that is, that the normal-metal island restores its equilibrium immediately after a tunneling event, we model the dynamics of the injector by the following stochastic equation on the number of excess electrons in the island $n$

\begin{equation}
\dfrac{d}{dt}p\left(n,t\right)=\sum_{n'\neq n}{\Gamma_{n'\rightarrow n}p\left(n',t\right)-\Gamma_ {n\rightarrow n'}p\left(n,t\right)}.
\label{markov}
\end{equation}
Here, $\Gamma_{i\rightarrow j}$ is the transition rate between individual charging states $i$ and $j$.

The explicit expression for the rates depends on the specific system and on the transition processes taken into account. Considering transitions in NIS junctions and only single and two-electron Andreev processes, we have
\begin{equation}
\Gamma_{n\rightarrow n\pm 1}\left(\varepsilon\right)=\dfrac{1}{e^2\rt}\int dE{n_\mr{s}\left(E\right)\left(1-f_\mr{N}\left(E+\varepsilon\right)\right)f_\mr{S}\left(E\right)} \label{e8}
\end{equation}
for single-electron tunnelling, and
\begin{equation}
\begin{split}
\Gamma_{n\rightarrow n\pm 2}\left(\varepsilon\right)&=\dfrac{\hbar\Delta^2}{16\pi e^4\rt^2\mathcal{N}}\int dEf_\mr{N}\left(E-\varepsilon/2\right)f_\mr{N}\left(-E-\varepsilon/2\right)\times\\
&\left|a\left(E+E_\mr{c}-i\delta/2\right)+a\left(-E+E_\mr{c}-i\delta/2\right)\right|^2 \label{e9}
\end{split}
\end{equation}
for Andreev tunnelling.
Here, $\varepsilon$ is the corresponding energy change in the tunneling event given by
\begin{align}
\varepsilon_{1e}\left(n\right)&=\pm 2E_\mr{c}\left(n-n_\mr{g}\pm 0.5\right) \label{e6}\\
\varepsilon_{2e}\left(n\right)&=\pm 4E_\mr{c}\left(n-n_\mr{g}\pm 1\right), \label{e7}
\end{align}
for single-electron $\left(1e\right)$ and two-electron $\left(2e\right)$ tunnelling, respectively.
In Eqs.~\eqref{e6} and \eqref{e7} $n$ is the initial island excess charge, $n_\mr{g}$ is the charge induced by the gate voltage, $E_\mr{c}$ is the charging energy and $+\,(-)$ designates tunnelling to (from) the island.

In Eqs.~\eqref{e8} and \eqref{e9}, $\Delta$ is the superconducting gap of the leads, $\rt$ is the junction normal-state tunnel resistance and $\mathcal{N}$ is the number of conduction channels which can be written as $A/A_{\mathrm{ch}}$ with $A$ being the junction area ($\approx 50\,\mr{nm}\times 60\,\mr{nm}$ for a typical device) and $A_{\mathrm{ch}}$ is the area of an individual channel (typically $30\,\mr{nm^2}$).
The term $\delta$ takes into account the energy of the intermediate (single-electron tunnelling) state which has a finite lifetime \cite{Averin2008} and therefore can be estimated as $\hbar\sum_\pm{\Gamma_{n\rightarrow n\pm 1}}$.
However, in our simulations and in the regime of interest the precise value does not have an impact on the calculated Andreev tunnelling rates~\cite{Maisi2014a}.
We set a reasonable value of $\delta/\Delta =10^{-5}$. Additionally, $f_\mr{N}$ is the Fermi-Dirac distribution of the normal-metal island, $f_\mr{S}$ is that for the superconducting lead and $n_s$ is the superconducting quasiparticle density of states given by 

\begin{equation}
n_\mr{s}\left(E\right)=\left|\mathfrak{Re}\left(\dfrac{E/\Delta+i\gamma}{\sqrt{\left(E/\Delta+i\gamma\right)^2-1}}\right)\right|,
\label{e11}
\end{equation}
here, $\gamma$ is the Dynes parameter which models subgap leaks and other junction non-idealities~\cite{Dynes1978,Pekola2010}.
Furthermore,
\begin{equation}
a\left(x\right)=\dfrac{1}{\sqrt{x^2-\Delta^2}}\ln\left(\dfrac{\Delta-x+\sqrt{x^2-\Delta^2}}{\Delta-x-\sqrt{x^2-\Delta^2}}\right).
\end{equation}

The probability $p$ obtained from Eq.~\eqref{markov} contains the whole information about the dynamics of the system.
To solve this problem one writes Eq.\eqref{markov} as

\begin{equation}
\dfrac{d}{dt}
\mb{p}\left(t\right)
= A\left(t\right)\mb{p}\left(t\right),
\label{markov_2}
\end{equation}
where $p_i\left(t\right)=p(n_i,t)$, with $n_i$ the $i$th charging state of the island, $A_{ii}=-\sum_{j\neq i}\Gamma_{i\rightarrow j}$ and $A_{ij}=\Gamma_{i\rightarrow j}$ for $i\neq j$.

However, one often is interested in calculating time-averaged transport quantities and not the vector of probabilities alone.
Of particular interest for the case of a frequency-to-power converter are the different moments of the transmitted energy to the superconductor $\left\langle E^m\right\rangle$.
To do this, one expands Eq.~\eqref{markov_2} so that
\begin{equation}
\dfrac{d}{dt}\begin{bmatrix}
\mb{p}\left(t\right) \\
\left\langle E^m\right\rangle_\mr{s}\left(t\right)
\end{bmatrix}
= \begin{bmatrix}
A\left(t\right) & \mb{0} \\
\mb{q}^{(m)}\left(t\right)^\mr{T} & 0
\end{bmatrix}
\begin{bmatrix}
\mb{p}\left(t\right) \\
\left\langle E^m\right\rangle_\mr{s}\left(t\right)
\end{bmatrix}.
\label{markov_extended}
\end{equation}
with $\mb{0}$ a $N\times 1$ vector of zeros with $N$ the total number of states considered. We have postulated that the time derivative of the state-averaged $m$th power of the transmitted energy holds $\frac{d}{dt}\left\langle E^m\right\rangle_\mr{s}=\mb{q}^{(m)}\cdot\mb{p}$, where $q_i^{(m)}=\dot{Q}^{(m)}_{i\rightarrow i+1}+\dot{Q}^{(m)}_{i\rightarrow i-1}$ such that
\begin{equation}
\dot{Q}^{(m)}_{n\rightarrow n\pm 1}\left(\varepsilon\right)=\dfrac{1}{e^2R_\mr{T}}\int{dE\left(E-\varepsilon\right)^m n_\mr{S}\left(E-\varepsilon\right)f_\mr{N}\left(E\right)\left[1-f_\mr{S}\left(E-\varepsilon\right)\right]}.
\label{stationary_moment}
\end{equation}
Again $\varepsilon$ is the related energy change from Eq.~\eqref{e6}.
The time dependences of the transition rates and of the vector $\mb{q}^{(m)}$ are implicit in their $\varepsilon$ dependence.
The time dependence in Eqs.~\eqref{e6} and \eqref{e7} is in the gate induced charge number $n_\mr{g}$ which follows the gate voltage $V_\mr{g}$ since $n_\mr{g}=C_\mr gV_\mr{g}/e$ with $C_\mr{g}$ the gate capacitance to the island.

The solutions of Eq.~\eqref{markov_extended} are given by
\begin{equation}
\begin{bmatrix}
\mb{p}\left(t\right) \\
\left\langle E\right\rangle\left(t\right)
\end{bmatrix}
=\exp{\left(\int_0^t{dt'\begin{bmatrix}
A\left(t'\right) & \mb{0} \\
\mb{q}\left(t'\right)^\mr{T} & 0
\end{bmatrix}}\right)}
\begin{bmatrix}
\mb{p}\left(0\right) \\
\left\langle E\right\rangle\left(0\right).
\end{bmatrix}
\end{equation}

In the specific operation of the frequency-to-power converter one assumes periodic boundary conditions on $\mb{p}$.
This is reasonable since the tunneling rates have the same periodicity as the driving signal.
As a consequence one can approximate the integral by discretizing the driving cycle of period $\tau$ in $l$ intervals of size $\delta t=\tau/l$.
At the end of the period the exponential is then
\begin{equation}
\tilde{U}(\tau)=\prod_{k=1}^l{\exp\left({\delta t\begin{bmatrix}
A\left(t_k\right) & \mb{0} \\
\mb{q}^{(m)}\left(t_k\right)^\mr{T} & 0
\end{bmatrix}}\right)},
\end{equation}
where $t_k=(k-1)\delta t$.

We decompose this propagator as.
\begin{equation}
\tilde{U}(\tau)=
\begin{bmatrix}
U\left(\tau\right) & \mb{0} \\
\mb{U}_q^{(m)\mr{T}}\left(\tau\right) & 0
\end{bmatrix}.
\end{equation}
Then, we enforce boundary conditions to get
\begin{equation}
\mb{p}\left(\tau\right)=\mb{p}\left(0\right)=U\left(\tau\right)\mb{p}\left(0\right).
\end{equation}
The calculation of $\mb{p}$ has been reduced to determining the eigenvector of $U\left(\tau\right)$ corresponding to the eigenvalue $1$.
Finally, we calculate the energy $m$th moment as $\left\langle E^m\right\rangle\left(\tau\right)=\mb{U}^{(m)}_q\left(\tau\right)\cdot\mb{p}(0)$.

\section{Parameters for the Heat Transport Calculations}

Table \ref{Param} presents the numerical constants used for modelling the heat transport across the SNS chain both in the full non-linear model and in the low temperature difference regime. The only free parameters remaining are initial power injected at the left end of S1 and the bath temperature. The former is fixed by the operation frequency $f$ of the converter as $\dot Q=2\Delta f$.

\begin{table}
\begin{tabular}{l | l}
Parameter & Value \\
\hline
\hline
$\ell_\mr n$ & $3.95\times 10^{-6}\,\mr m$ \\
$\ell_1$ & $7.00\times 10^{-7} \,\mr m$ \\
$\ell_2$ & $6.00\times 10^{-6}\,\mr m$ \\
$A_\mr N$ & $3.60\times 10^{-15}\,\mr m^2$ \\
$A_1=A_2$ & $6.00\times 10^{-15}\,\mr m^2$ \\
$\rho$ & $3.10\times 10^{-8}\,\mr{\Omega m}$ \\
$\rho_\mr N$ & $1.69\times 10^{-8}\,\mr{\Omega m}$ \\
$\Sigma_\mr S$ & $1.80\times 10^{9}\,\mr{WK^{-5}m^{-3}}$ \\
$\Sigma_\mr N$ & $7.00\times 10^{9}\,\mr{WK^{-5}m^{-3}}$ \\
$\Delta$ & $2.00\times 10^{-4}\,\mr{eV}$ \\
\hline
\end{tabular}
\caption{Parameters used in the calculations of heat transport. $\rho_\mr N$ is the normal-metal resistivity, the remaining symbols are defined in the main text.}
\label{Param}
\end{table}